\documentclass[11pt]{article}
\usepackage{amsfonts}
\usepackage{amsmath}
\usepackage{natbib}
\usepackage{hyperref}
\usepackage{graphicx}
\usepackage[margin=4.0cm]{geometry}
\usepackage{multirow}
\usepackage{array}
\usepackage{pdfpages}

\DeclareMathOperator*{\argmax}{arg\,max}
\DeclareMathOperator*{\range}{\mathrm{range}}
\newtheorem{myprop}{Proposition}
\def\materials{\vskip6pt\bgroup
\noindent {Materials and Methods}\\[2pt]
}
\def\acknowledgments{\vskip14pt\bgroup\footnotesize\baselineskip 8pt
\noindent{\bf ACKNOWLEDGMENTS.} \relax\ignorespaces}
\def\endacknowledgments{\vskip1sp\egroup}

\def\endmaterials{\vskip1sp\egroup}

\begin{document}

\title{Inequality, unrest and the failure of states: a qualitative modelling framework}

\author{Daniel John Lawson\footnote{Heilbronn Institute for Mathematical Research, University of Bristol, Bristol, BS8 1TW, United Kingdom} \and
Neeraj Oak\footnote{Department of Engineering Mathematics, University of Bristol, Bristol, BS8 1TW, United Kingdom}}

\maketitle
\begin{abstract}
An explanation for the political processes leading to the sudden collapse of empires and states 
would be useful for understanding both historical and contemporary political events.
We seek a general description of state collapse spanning eras and cultures, from small kingdoms to continental empires, drawing on a suitably diverse range of historical sources. 
Our aim is to provide an accessible verbal hypothesis that bridges the gap between mathematical and social methodology. 
We use game-theory to determine whether factions within a state will 
accept the political status quo, or wish to better their circumstances through costly rebellion.
In lieu of precise data we verify our model using sensitivity analysis.
We find that a small amount of dissatisfaction is typically harmless, but can trigger sudden collapse when there is a sufficient buildup of political inequality.
Contrary to intuition, a state is predicted to be least stable when its leadership is at the height of its political power and thus most able to exert its influence through external warfare, lavish expense or autocratic decree.

\end{abstract}




\section{Cycles and collapses in history}

History has witnessed the rise and fall of countless empires, dynasties and regimes. What governs these apparently inevitable processes has been discussed across the eras \cite{breisach2007historiography}. 
Whilst growth and power seem naturally self-reinforcing, reversal into decline or collapse has impacted every state and culture not present today.
Further, the fate of a nation is often tied closely to the fate of its leading class; the sudden collapse of one usually leads to a similar collapse of the other \cite{Boyle1968,Scales1993}.

Within a state, influence and power are often distributed unequally. Political change is affected by many factors including visible achievements and failures, deliberate manipulation, accidents of fate and external forces. Historically, stable political states can enjoy long periods of relative growth and internal stability during which the leading class can gain a larger and larger share of the wealth and resource \cite{gibbon1776history,Scullard2010}. However, the process of mounting inequality has clearly not continued forever.

Power may also change rapidly, and with great impact on the fate of apparently stable states. Whilst we do not apply our model to contemporary conflict, the clarity provided by modern media during the Arab Spring of 2011 \cite{johnstone2011global,campante2012arab} illustrates the lack of simplicity in these transitions. 
In many cases, rebellion operated without a unified name or organisation long before any form of leadership emerged (for example, in Libya \cite{gause2011middle}), 
signifying a decentralised process.

We are interested in why social disorder appears rapidly from an apparently stable state.  Is there a generality describing when dissident movements will receive support and when they will be ignored?  Actual success of rebellion movements means acquiring military power, which is strongly dependent on technology and social structure.  Those with the military power may join the rebellion if it is in their interests to do so. During peace this may seem implausible, but the toll of rebellion may rapidly change the situation.

Collapse events have been linked to environmental factors such as local or global climate change \cite{cullen2000climate,weiss2001drives,Zhang2007,buckley2010climate,Dugmore2012} and long term degradation of resource \cite{diamond2005collapse} (although there is still controversy, e.g. \cite{marohasy2005australia}). However, the environment alone is unlikely to provide a general explanation for collapse. Disturbance does not affect all societies equally; for example, Sassanid Persia thrived during periods in which the neighbouring Roman empire experienced agricultural decline \cite{McCormick2012}. Many collapse events occur in the absence of environmental pressure \cite{butzer2012collapse}, with external conquest, internal conflict, or poor social, political and economic institutions playing a greater role instead.  Our model describes the dynamical process behind the social conditions that make unrest more likely in the presence of external stresses.

Our model is complementary to other theories of collapse \cite{tainter1990collapse} by providing a game-theory or economic explanation for social assumptions.
Collins \cite{Collins1978,Collins1986} emphasises the importance of areas at the fringe of empires, so called `marchlands', which tend to be the incubators of new regimes or polities. 
The thirteenth century author Ibn Khaldun \cite{Khaldun1958} describes a concept he calls `asabiya' or `group feeling' in which loyalties are nested within a state.  The metaethnic frontier theory of Turchin \cite[p.50-77]{TurchinHistoricalDynamics} combines these hypotheses.  As we predict that power equality can lead to stability, the most cohesive states should emerge from marchlands and tight-knit groups with high asibiya.

We join a recent trend of providing mathematical models for historical hypotheses (some excellent examples are \cite[]{Turchin2003,knappett2008modelling,currie2010rise,baggaley2012bayesian}).  
Whilst some general principles of civil conflict and disunity are understood \cite{AghionWilliamson1998,Collier2000} without the need for modelling, 
mathematics provides formal reasoning that aids in generalisation. 
A mathematical theory of collapse is a first step towards a statistically sound, 
data-driven comparison between hypotheses (a feat we do not attempt here).  
Our model is too general to be the full explanation for any specific scenario, 
so we consider a wide range of documented collapse events that contain qualitative similarities
without claims about the important factors in any given situation.
Conceptually the model is qualitative and robustly explored by considering numerous precise instantiations,
which acts as a sensitivity analysis \cite{saltelli2000sensitivity} helpful for supporting (but not confirming) conclusions from non-quantitative data.

\section{A qualitative model of collapse}
\label{sec:qualitative}

\begin{figure}[!ht]
\begin{centering}
\includegraphics[clip,width=0.97\textwidth,angle=0]{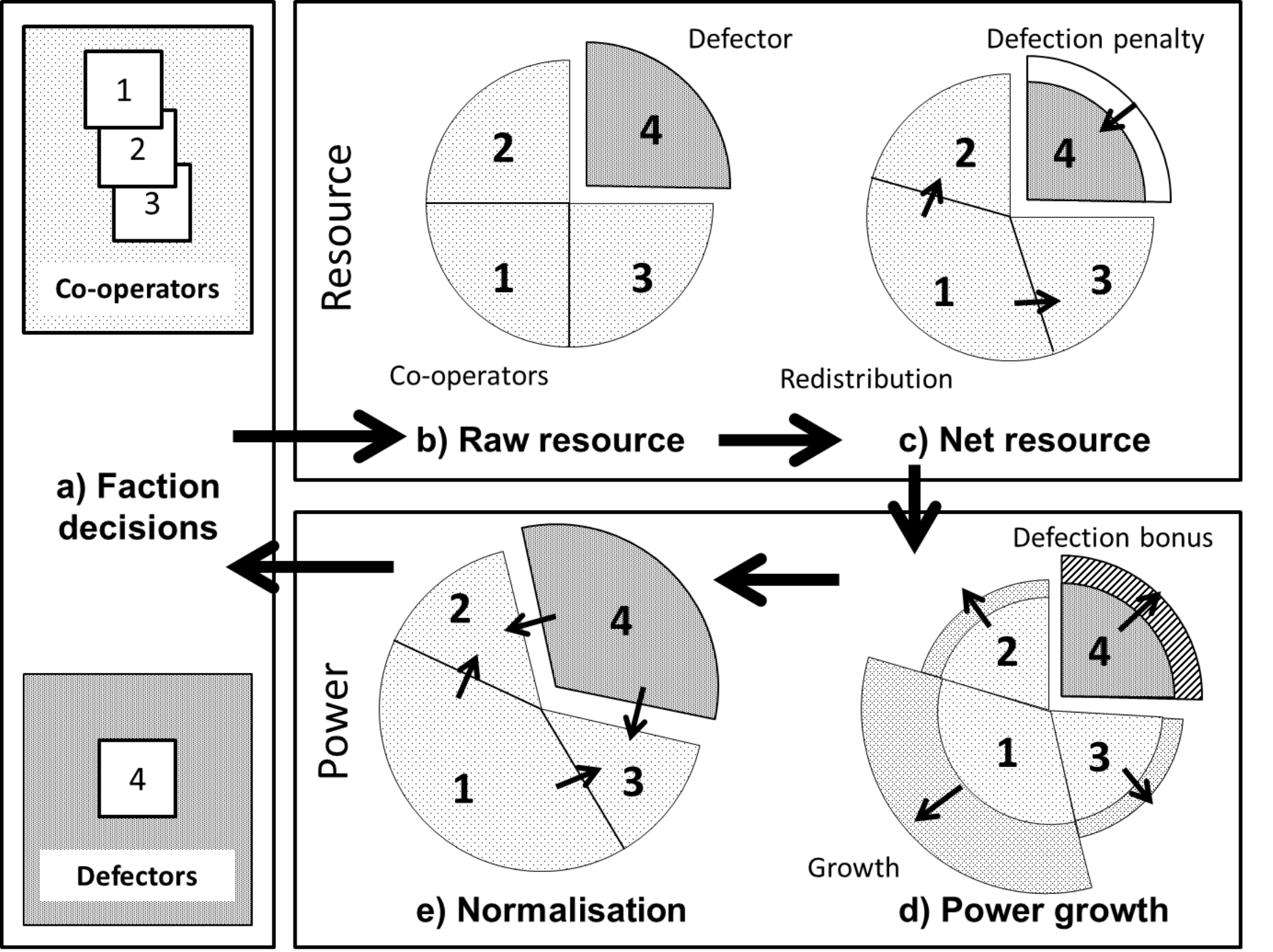}
\caption{Qualitative model.  Factions (a) decide whether to cooperate or defect. Then  (b) raw resource is collected, which (c) is either reduced (for defectors) or redistributed according to power (for cooperators). Power grows (d) according to resource, with a defection bonus, and (e) is normalised. This effectively reduces power for some and increases it for others, potentially changing their behaviour next round.
\label{fig:qualitativemodel}}
\end{centering}
\end{figure}

Consider a number of actors playing a repeated public goods game, in which cooperators enter their resource into a public pool to be redistributed according to influence, which changes over time.  Defectors obtain lower mean payoff but are not subject to redistribution. The game dynamics  (Fig. \ref{fig:qualitativemodel}) draw on three vital qualitative assumptions:
\begin{enumerate}
\item Inequality of influence and hence resource will increase (on average) over time when actors cooperate.
\item Defecting produces an overall cost, reducing resource for the defector and reducing the public goods for the cooperators.
\item Defecting decreases the future influence inequality.
\end{enumerate}
Cooperating means obeying the rules of a political system designed to prevent costly conflict between the actors. Within the system, political influence tends to accumulate with those that have the most resource, leading to increasing inequality.  By `defecting' from the political system, actors pay a cost but increase their political standing.

We investigate how these assumptions lead to coordinated activity such that actors cooperate periodically, and defect en-masse.  This dynamic may provide an explanation for the long-term difficulties experienced by many co-operation systems, including the disintegration of powerful nation-states and empires over the course of history.

\section{Qualitative trends in history}
\label{sec:history}

In this section we outline some examples of qualitative features that are consistent with our model.

\subsection{The ubiquity of collapse}
The phenomenon of collapse has occurred across diverse world cultures throughout history, and has affected polities of all sizes. 
Here we describe cases where a state able to exert significant power has experienced collapse or unrest.
Such states have strong leadership relative to their subjects, evidenced by a) the successful imposition of will, or b) the ability to expend resource in prolonged offensive wars, extensive building programs, etc.

The reign of Amenhotep IV (Akhenaton) of Egypt is an excellent example of unrest stemming from centralised power, demonstrated by autocratic decree and lavish expense \cite[p.204-205]{Cambridge1924}. He decorated his capital and empire with sun temples and other paraphanalia of a new religion, which led to unrest and internal disorganisation \cite[p.207]{Cambridge1924}.

The early reign of the Achaemenid Persian emperor Xerxes, following that of his father Darius, was beset by internal rebellion in Babylonia and Egypt. Darius' reign involved extensive military activity, and while Xerxes was eventually able to return to the offensive, his reign was necessarily more passive and defensive \cite[p.78]{Cambridge1988}.

The division of Alexander the Great's empire can be thought of as a collapse, as it resulted in violent upheaval \cite[p30-69]{Waterfield2011},\cite[p3]{Bickerman1983}. The successor states or Diadochi engaged in extensive warfare over the following decades, and many subdivided further \cite[p7]{Bickerman1983}. 

Many examples can be taken from the late Roman republic; the Social (or Marsic) War \cite[p.68-70]{Scullard2010}, as well as three Servile Wars \cite[p.95-96]{Scullard2010} were fought by Rome during or immediately after periods of expansion abroad. The stated grievance in these cases was explicitly inequality; the slaves wished to be elevated from their abject position in society, whilst the Socii demanded an end to their second-class status within the Republic. Caesar's civil war \cite[p. xxvii-xxx]{CaesarCivilWar} also took place immediately after a period of offensive foreign campaigns, notably the annexation of Gaul. Caesar also explicitly touted the grievance of inequitable distribution of wealth and power \cite[p.76-77]{CaesarCivilWar}.

The reign of the Roman emperor Domitian was lambasted by writers of his time as being tyrannical and autocratic \cite{Fritz1957}. Contemporary detractors claimed that he ignored tradition, executed senators who opposed him and openly asserted primacy over the senate \cite[p.193-198]{Jones1992}. After his assassination, the new emperor Nerva had a short and impotent tenure, due to revolts by the military. The indulgence and autocracy of the Roman emperor Commodus \cite[p.64-79]{Grant1996} is often described as the beginning of the decline of the Roman empire \cite{gibbon1776history}, leading to a protracted civil war. Yet the Nerva-Antonine dynasty that preceded Commodus was amongst the most successful periods in Roman history. The remnants of the Roman empire, centred on Byzantium, exhibited periodic instability well into the Middle Ages \cite[p258-259]{Auzepy2008}. 

Immediately after the adoption of Islam, Arab armies expanded their domain westwards towards Morocco, east into Persia and as far north as France. The empire collapsed in the mid-8th century, and the successor states suffered further internal conflict and dissolution in the following centuries \cite[p1-10]{Scales1993}.

The Angkor or Khmer civilisation of Indochina experienced sharp episodes of civil war, notably prior to the accession of Suryavarman I \cite[p112]{Coe2003}. Angkor provides an excellent example of a state whose power rested on appropriation from vassals; indeed, much of Jayavarman II's power depended on a rice surplus \cite[p89-90]{Higham2004}.

Richard I of England undertook extensive conflict in France and the Levant. His reign was followed by that of John I, in which concessions were made to the nobility in response to mounting unrest \cite{Hollister1961}.

In the wake of their initial conquests, the Mongol empire divided, at times violently, into smaller, culturally heterogeneous polities \cite[p340-355]{Boyle1968}. 

The 1905 \cite{Ascher1988} and February 1917 \cite{Polunov2005} revolutions in Russia are often attributed to the inequitable distribution of wealth and power\footnote{The term `autocrat' was part of the official title of the Tsars.} between the ruling classes and the majority of the population \cite{Trotsky1972}.

\subsection{Trends to conglomeration}

Although there are counter-examples, the tendency towards conglomeration is relatively ubiquitous. We note the latifundia of ancient Rome \cite{gibbon1776history}, the extensive provincial landholdings of the late Sassanid empire \cite{Eisenstadt1964}, and Feudal Europe \cite{BlaydesChaney2013}. Mercantile quasi-states such as the Hanseatic League \cite[p186]{Dollinger1964}, Italian merchant republics \cite[p212-229]{Hunt1994} and European colonial enterprises \cite[p250-254]{Gardner1971} also exhibit this behaviour. 

\subsection{Cascading civil unrest}

Uncoordinated defections often produce their own momentum; as more factions choose to leave a state, they impact upon the perceived legitimacy of that state and encourage further defections. Several previous examples exhibit this behaviour \cite{Waterfield2011,Bickerman1983,Auzepy2008,Laiou2008,Herrin2007,Gregory2005,Treadgold1997}.


\subsection{Revolt through desperation}

A common instigator of internal conflict is the inability of a faction to hold political power or influence by peaceful means. This occurs when a faction perceives that it receives an inadequate degree of power, or that it's power is being eroded.  The perception of inequity may encourage the use of violence in order to redress the apparent imbalance. Excellent examples of this appear in the late Republican period of Rome \cite[p.405-446,505-543,619-674]{DillonGarland2005} ; of particular note is the career of Catiline \cite[p.142]{Durant3} . The repeated, and often unsuccessful, peasant revolts of medieval Europe also exhibit many of these features \cite{Bernard1964}.


\section{Basic Model: state formation and collapse}
\label{sec:model1}

The model takes the form of an iterated multiplayer game. Consider $N$ autonomous factions (i.e. actors), who may be individuals or groups with similar enough motivations to act coherently, competing via the model in Figure \ref{fig:qualitativemodel} (defined precisely in Methods).  Each faction starts with an equal amount of raw resource and chooses to defect or cooperate in order to obtain the highest net resource, assuming that other factions do not change their decisions. Net resource depends on all factions' actions and power. Power is then updated\footnote{Power is treated a zero-sum game, although this is a simplification of reality \cite{baldwin1979power}.} based on the resources obtained and the decisions made, increasing most when defecting or when a large amount of resource is obtained.  The process is then repeated using the new distribution of power and decisions.

\begin{figure}[!pht]
\begin{centering}
\includegraphics[clip,width=1.0\textwidth,angle=0]{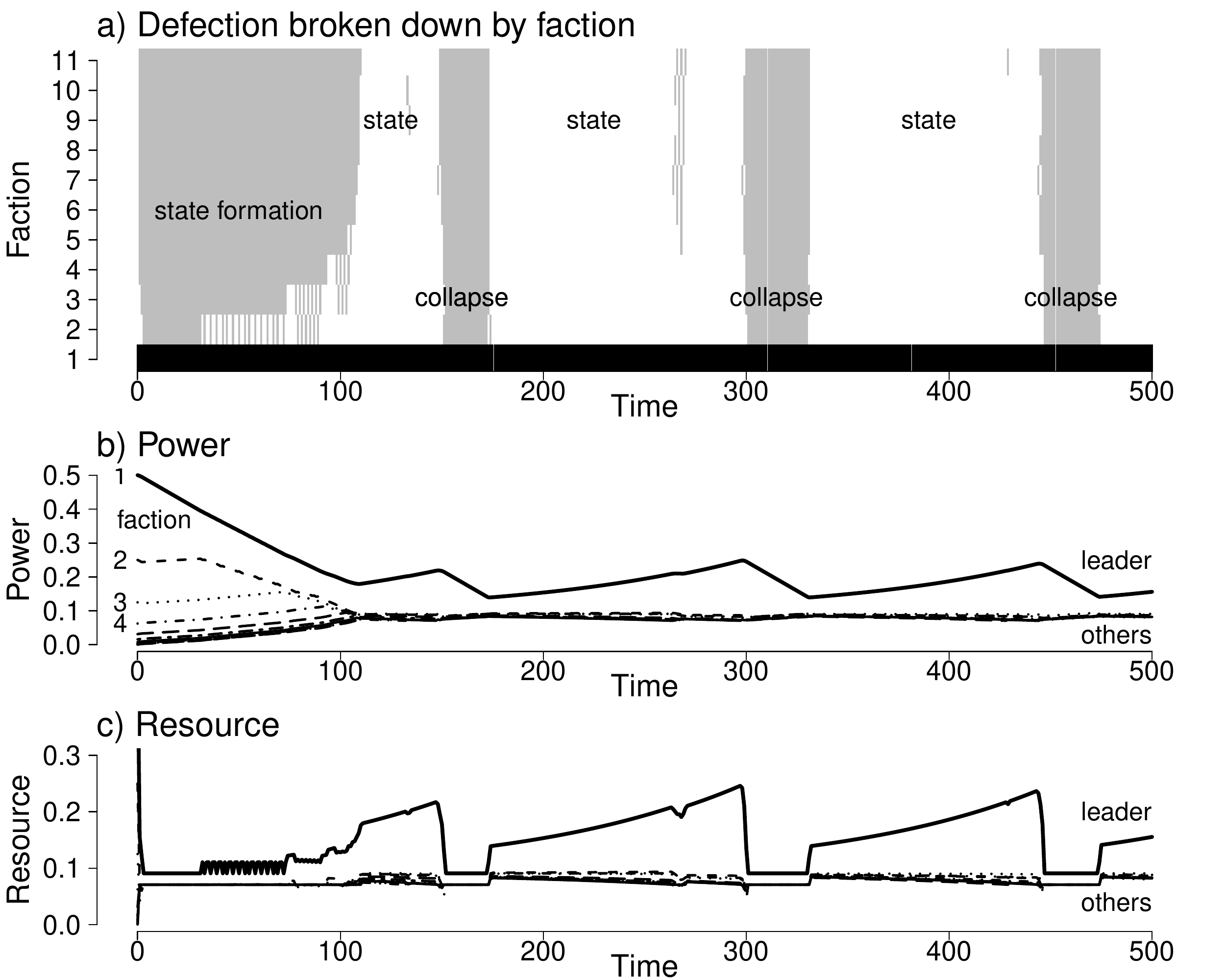}
\caption{Characteristic behaviour of our model.  a) Defection behaviour with state formation and collapse. Defection is shown in grey, cooperation in white, and the leading faction in black (which always cooperates).  b) The power of factions over time. c) The resource of factions over time. The power and resource of the non-leader factions converge, with the result that periodic coordination and defection periods occur. (Parameters: $\rho=0.2$, $w=0.02$, $N=11$ and $\mu=0.01$.) 
\label{fig:Model1Example}}
\end{centering}
\end{figure}

The model produces periods of widespread cooperation and collapse, as shown in Fig. \ref{fig:Model1Example}. During state formation, the power of non-leader factions equalises around the point at which cooperation becomes viable. 
Once cooperation is established, defection occasionally occurs in isolation as redistributed resource falls below a threshold.  Rarely, but periodically, enough factions are sufficiently weak so that the defection of the preceding faction changes their own best choice, leading to a defection cascade.  Once the political landscape has equalised sufficiently, a corresponding cooperation cascade occurs nearly as rapidly.  Cooperation and defection periods occur with a predictable timescale leading to `spontaneous' periodic behaviour.

\begin{figure}[!phtb]
\begin{centering}
\includegraphics[clip,width=1.0\textwidth,angle=0]{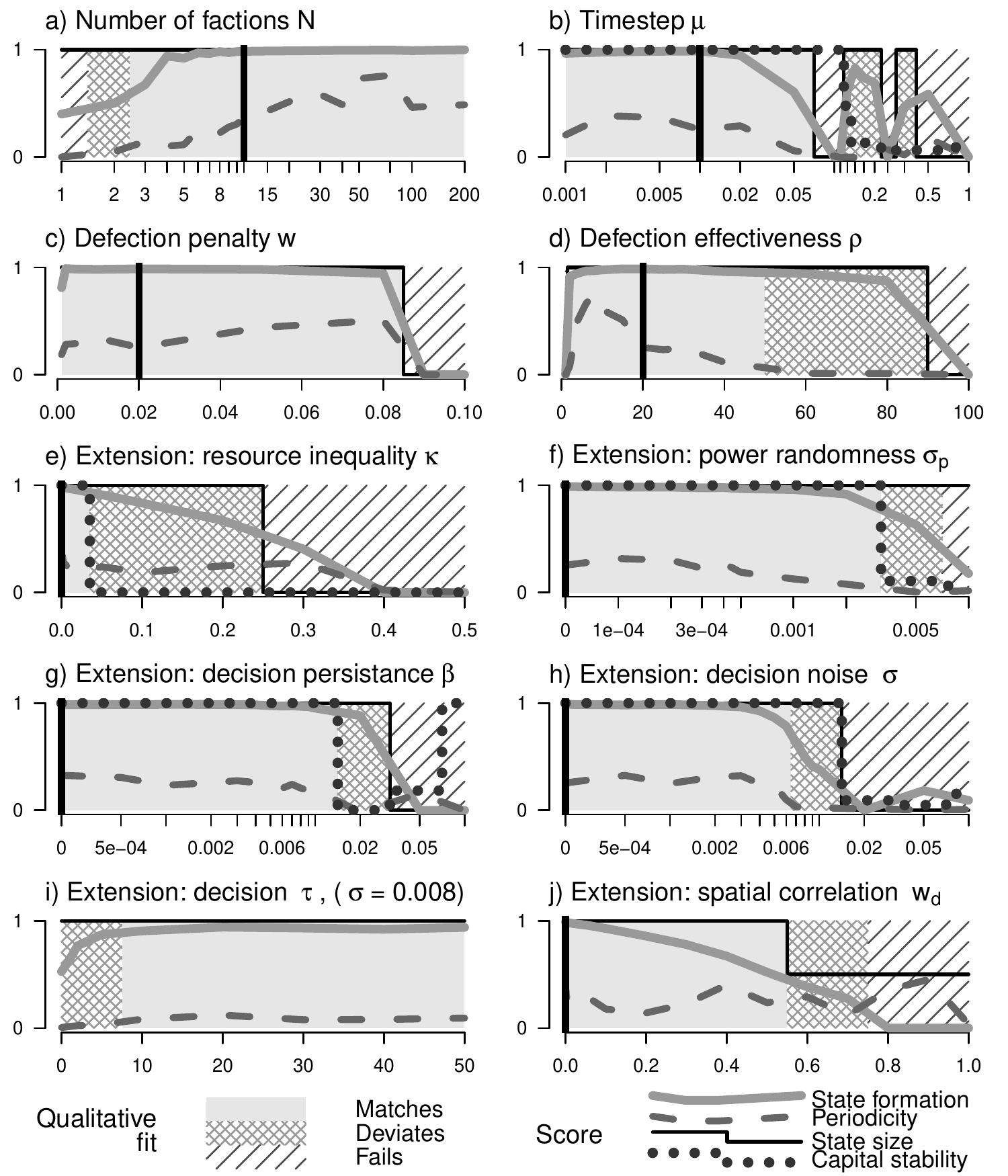}
\caption{Effect of parameters/model extensions on the qualitative dynamics. The plots are shaded to show whether model qualitatively behaves as Figure \ref{fig:Model1Example}. The model either matches (solid), deviates (dense hatching) or fails (thin shading). The qualitative fit is based on quantitative scores (see Methods). Firstly, `State formation' ($S_f$) is high when states are large and collapse rapidly to few factions. Secondly, `Periodicity' ($S_p$) is high if there is periodic predictability to decisions. Thirdly, `State size' ($S_s$) is high if state formation and collapse affect all factions. Finally, `Capital stability' ($S_c$) is high if the leading faction does not change from the initial leader (relevant only for plots e--h).  The qualitative model is matched if $S_f>0.5$, $S_p>0.05$, $S_s=1$ and $S_c=1$. It deviates if $S_p<0.05$, $S_s=0.5$ or $S_c=0$.  Otherwise the qualitative model fails.  Also shown (where possible) is the parameter value from Fig. \ref{fig:Model1Example} (vertical line).
\label{fig:allsummaries}}
\end{centering}
\end{figure}

How reliable is the model?  It has four parameters: the number of factions $N$, time discretisation $\mu$, defection resource penalty $w$, and defection power gain $\rho$.
Fig. \ref{fig:allsummaries}a-d highlights the parameter regions for which the model behaves as Fig. \ref{fig:Model1Example}. Although $N$ has many important effects (Section S1), 
all values of $N>2$ match the qualitative model. Similarly, all theoretically valid values of the defection penalty $w$ also match (Section S1.3). 
Small timesteps $\mu<0.05$ act as a timescale, but large $\mu$ or rebellion effectiveness $\rho$ prevents periodic behaviour due to `intrinsic' noise from the discretisation of time (see below).

\subsection{Features of the Basic Model}
\label{subsec:reasoningmodel1}

We can take the continuous time limit of our model, which removes intrinsic noise due to discretisation. We can also take the continuous faction limit, which leads to a Partial Differential Equation model.  These models (Section S1) 
 are not readily solvable but do allow us to understand why our model behaves as it does.

During cooperation, power departs exponentially from $R_0$. Defection occurs because:
\begin{enumerate}
\item A defection always makes cooperators worse off;
\item Later defections have a greater impact than early defections, making a cascade more likely as more factions defect;
\item Failed defection cascades erode the power of weaker non-defecting factions most, helping future cascades to succeed.
\end{enumerate}

The dynamics follow 4 distinct phases that repeat in a cycle:
\begin{enumerate}
\item Cooperation: Power becomes concentrated in the leading factions.  Weaker factions may defect in an uncoordinated manner. 
\item Collapse: Defectors coordinate into a cascade when the cumulative power distribution is everywhere above a threshold.
\item Defection: Defection continues until power becomes sufficiently diffuse to permit cooperation.  The strongest factions may cooperate first in failed state formation attempts.
\item Recovery: A cooperation cascade occurs in much the same way as the defection cascade, when the cumulative power distribution is everywhere below a (complex) threshold.
\end{enumerate}

Additionally, we obtain a bound on the duration of cooperation and defection periods by allowing all non-leading factions to behave identically. In this case we can obtain closed-form expressions for the duration of cooperation and defection phases.  The initial conditions can be very important in determining how close the bound is, from which we conjecture that this model has no general analytic solution, although bounds can be found and special cases solved.

\section{Extensions: A model sensitivity analysis}
\label{sec:theory}

When performing parameter inference using quantitative data, a minimum requirement is to assess how robustly the parameters are inferred via a sensitivity analysis \cite{saltelli2000sensitivity}. Here we are instead trying to infer that some qualitative features were created by a general class of model. We attempt to understand the qualitative model space using a model-level sensitivity analysis. 

\subsection{Unequal resource distribution}
Resource is distributed unevenly in practice, which we model by replacing $R^0$ with $R^0_i\propto \exp(\kappa i)$ for faction $i$. $\kappa>0$ means that initially powerful factions have less resource. Fig. \ref{fig:allsummaries}e shows that a resource-weak leader can either persist or be usurped. Periodicity and collapse events persist, and further, Section S2.1 and Fig. S1 
show that a resource-weak leader results in reduced average conflict. 

\subsection{Uncertain outcomes}
The political power process is contingent on events outside of complete control of faction leaders. We model this by adding noise (normal, with mean $\sigma_p$; see Section S2.1) 
to the obtained power change before normalisation.  Fig. \ref{fig:allsummaries}f shows that small levels of noise do not effect the qualitative behaviour.  Moderate levels lead to a leader turnover and loss of periodicity, whilst high levels prevent both coordination of both state formation and collapse (Fig. S2).

\subsection{Biased decision making and random choices}
People are not naive resource optimisers. Decisions may be biased, hard to change, poorly calculated, made with respect to longer term goals, or otherwise unobserved features.

Complex decisions can be allowed for by introducing a \emph{random function} $\eta_i(t)$ for the decision threshold of each faction. We include `persistence' via a bias $\beta$ towards the previous action, and random fluctuations in whether to favour defection or cooperation via a \emph{Gaussian Process} \cite{rasmussen2006gaussian} (Section S2.3). 
This is determined by two parameters: the magnitude of the fluctuations $\sigma$ and their correlation over time $\tau$. 
To interpret $\tau$, factions are effectively making `new random decisions' every $O(\tau)$ time units.  If $\tau \ll 1$, then decisions appear `noisy', and if $\tau$ is very large, each faction will appear (randomly) biased.

Fig. \ref{fig:allsummaries}g and Fig. S3 
show that a range of $\beta$ has no qualitative effect, whilst moderate values lead to leader turnover and a lack of periodicity.  Fig. \ref{fig:allsummaries}h-i demonstrate that small to moderate levels of noise (Fig. S4) 
don't effect the dynamics, and further, when decisions are more correlated in time  (Fig. S5) 
then state formation is more stable to high decision noise. This happens  
because power has time to equilibrate around the random choice of decision boundary.

\subsection{Spatial structure}
Some political scenarios are best described with a spatial model.  For example, factions may be local leaders of villages, or semi-autonomous regions of a larger state.  We replace $w$ by $w_i=w^* \exp\left(-|x_i - x_C| w_d/N \right)$ (Section S2.4), 
i.e. both the resource penalty for defection, and the political gains from doing so, decay with distance from the capital (leading faction).  The average $\mathbb{E}(w_i)=w$, i.e. is unchanged, and distance is calculated on a ring (so factions $1$ and $N$ are neighbours).

The spatial model (Fig. \ref{fig:allsummaries}j)  allows for a variety of different scenarios.  The state grows from the capital (Fig. S6) 
and collapses as in the non-spatial model.  Collapse may be from the outside in, or the inside out.  There may be a well defined maximum spatial extent (hatched region of Fig. \ref{fig:allsummaries}j).

\subsection{Modified intrinsic noise}
We chose to define Model 1 as an iterated game, which has consequences for the way that noise enters the system.  Although the basic model is deterministic, the discretisation of time can produce `chaotic' dynamics (e.g. large $\mu$ in Figure \ref{fig:allsummaries}b) as small variations in the value of political power have large effects.  Is this `intrinsic noise' important for the dynamics?

To address this issue we constructed a modified version of the model in which only a single faction makes a decision at a time (using the Gillespie Algorithm \cite{gillespie1977exact}) with an average timestep of $\mu/N$ (Section S2.5). 
Fig. S7 
compares this model with the basic model and shows that there is no qualitative change.  Additionally, the continuous time version of the model (Section S1) 
matches the qualitative data. We view these issues as `modelling degrees of freedom' and only consider model behaviours that are present for all choices.

\subsection{Non-uniform defection penalty} 
If the penalty for defection decreases with the number of defectors, both defection during cooperation and cooperation during defection are harder. This makes the phenomenon of periodic collapse more likely to occur, as we show numerically (Section S2.6 and Fig. S8). 
Leader replacement is also easier in the presence of intrinsic (or extrinsic) noise.

\subsection{Non-linear relationship between power and resource} 
Power and resource are simply related in our model. However, we find that a family of non-linear functions do not effect the qualitative dynamics (Section S2.7). 
Since we can map resource levels to a decision boundary in the power distribution, we conjecture that most `reasonable' increasing functions will demonstrate collapse.

\section{Game Theory perspective}
\label{subsec:gametheory}

We have not permitted factions to consider politics when making decisions.  Do societies still collapse when longer term strategies can be employed? A little game-theory analysis shows that the main phenomena persist, and further, that the game has interesting behaviours of its own.

The resource payoff in our model takes the form of a simple iterated (multiplayer) game.  Consider the case where there are two factions, $i$ and $j$ (with $p_j> 1/2 > p_i$ and $p_i+p_j=1$) having payoff structure:
\begin{equation*}
\begin{array}{c|r l r l}
\mathrm{Payoff\, for}\,\, i\, | \, j & \multicolumn{2}{c}{j \mathrm{\,defects} }& \multicolumn{2}{c} {j \mathrm{\,cooperates}} \\ \hline
i \mathrm{\,defects} & \multicolumn{1}{c|}{1/2-w} &1/2-w & \multicolumn{1}{c|}{1/2-w} &1/2 \\
i \mathrm{\,cooperates} & \multicolumn{1}{c|}{1/2} &1/2-w & \multicolumn{1}{c|}{p_i} &p_j \end{array} \label{eqn:payoff1}
\end{equation*}
Until now we have assumed that factions are simple resource maximisers. Whatever $i$ does, $j$ always obtains more resource from cooperation. Since $i$ knows this it should cooperate only if $p_i>1/2-w$.  In this circumstance $p_i$ increases when $i$ defects, and decreases when it cooperates.  In Model 1a where decisions are continuous, $i$ changes action around the decision boundary and obtains payoff $1/2-w$ whether cooperating or defecting.

However, $i$ could attempt to maximise its payoff over all time. There is no reason for $i$ not to defect for the power benefit, since only the defection payoff is obtained on average. If $i$ were to defect until power is equalised, then it would enjoy a long period of high resource until power became uneven again. The payoff from becoming the leading faction is even higher.  

Should $i$ agree to cooperate if $p_i<p_j$?  If not, and $j$ uses the same reasoning, then both players will on average get the `greedy' payoff $r_g=1/2-w/2$ (as they have to defect half of the time).  If either were willing to take the lower cooperation payoff they would get more over time. Proposition \ref{prop:gametheory} shows that there is a strategy which maximises the long-term resource payoff:
\begin{myprop}\label{prop:gametheory}
For Model 1a (the continuous time model), there exists a defection strategy defined by a power lower bound $p_{min}=p_{max}-a$ for a given upper bound $p_{max}\le 1/2-\delta$ with $a>\delta>0$, which when both players use it the resource obtained is $r^*_{high}(a,\delta) > r_{low}^*(a,\delta) > r_g$ for the stronger and weaker players respectively. (For proof, see Methods).
\end{myprop}

The existence of this longer term strategy leads to an interesting extension. We now allow $i$ and $j$ to use two potential strategies in a long-term meta-game, which both dominate the short term strategies.  Passive players use the strategy from Proposition \ref{prop:gametheory}, and aggressive players cooperate only as the dominant faction.  We consider the payoff averaged over many periodic cycles, assuming that during each state formation the player with the initially higher power is chosen randomly. Therefore a passive player always ends up with lower power than an aggressive player, two aggressive players obtain the `greedy' payoff $r_g$, and two passive players share leadership over time. The average payoff matrix is:
\begin{equation*}
\begin{array}{c|r l r l}
\mathrm{Payoff\, for}\,\, i\, | \, j & \multicolumn{2}{c}{j \mathrm{\,aggressive} }& \multicolumn{2}{c} {j \mathrm{\,passive}} \\ \hline
i \mathrm{\,aggressive} & 
\multicolumn{1}{c|}{r_g} &r_g & \multicolumn{1}{c|}{r_{high}^*} & r_{low}^*\\
i \mathrm{\,passive} & 
 \multicolumn{1}{c|}{r_{low}^* } & r_{high}^* & \multicolumn{1}{c|}{\frac{r^*_{high}+r^*_{low}}{2} } &\frac{r^*_{high}+r^*_{low}}{2}  \end{array} 
\end{equation*}
As $r_g<r_{low}^*<r^*_{high}$, this is a form of the prisoners' dilemma.  We note an analogy of the `passive' strategy with democratic parties sharing power over time, as opposed to corrupt or autocratic systems in which this is impossible.

The case of three or more factions can be understood analogously. Optimally, weaker factions will act together to remove power from the leader, and such coordination follows naturally from the `passive' strategy. Consider that factions cannot solve for the optimal thresholds $a_i$ but can choose them empirically. Cooperation first occurs when all factions have equal power (to within $\delta$).  Each defection now occurs at the factions' chosen threshold. As in the simple model, defections reduce cooperators' resource leading to defection cascades.  Aggressive factions still do not cooperate unless they are the leader and will not take part in the state.  We therefore conjecture that any number of short-term resource maximisers, and/or `passive' long-term strategists, can form states that experience dramatic collapse events.

\section{Discussion}
\label{sec:discussion}

Proper empirical validation requires a wide and unbiased range of data sources. Unfortunately, historical data has not been curated into a quantitative form en-masse as evidence is patchy and inherently qualitative.  We hope recent efforts towards quantification \cite{Turchin2012} will allow our model and others to be subjected to trial by data.  Until then, all general models have been selectively and qualitatively validated and are simply plausible explanations for collapse events.  More positively, many of the models mentioned above share our qualitative assumptions and could cite the same historical case studies in their defence.

Collapse is inevitable in our model but real states may use unmodelled options. For example, they can attempt to eliminate rebelling factions before the collapse cascade begins, or maintain an equitable resource distribution by deliberately limiting their own power.  The efficacy of such strategies is not considered here.

Conflict is considered only as a reduction of resource in our model and so force is only used to impose sanctions.  This may seem unrealistic when many civil wars are put down violently.  
Whilst we might consider that the elimination of defecting factions could create new dissatisfaction, this has not been modelled.
Our hypotheses as described best represent political systems that dissuade escalating warfare.  Historical examples are coalitions of city states (e.g. Ancient Greece), or feudal lords (within a medieval European country), who have strong cultural bonds and may have previously existed within a larger state. 

We have made a significant effort to legitimatise the use of a utility function for faction behaviour, by incorporating random time varying functions into the decision process.  Many factors influence decision making, from alternative goals to incomplete knowledge, without a need to address whether the choices are rational.  The qualitative assumption here is that the resource difference between defection and cooperation will correlate with the choice a faction makes.

A final, but vitally important point is that our model (like all models of complex systems) may be incorrect in specific details and incomplete in general.  Such quantification of hypotheses  is still helpful as it makes it possible in principle to draw statistical comparisons between explanations for collapse. We remain hopeful that their relative contributions can be scientifically assessed to further our understanding of political history.

\begin{materials}\label{sec:materials}


\subsection{Mathematical Model}
The decision to defect $D_i(t+1)=1$ if $\eta_i(t+1)<0$ and $D_i(t+1)=0$ (cooperation) otherwise, where:
\begin{equation}
\eta_i(t+1) = R_i(t+1 | D_i=0, \mathbf{D_{-i}})
 - R_i(t+1 | D_i=1, \mathbf{D_{-i}}),
\end{equation}
and $R_i(t+1| D_i)$ is the predicted resource obtained by action $D_i$ (assuming all other factions do not change). Cooperators pool and redistribute resource, 
$R_i(t+1 | D_i=0) = R^0_C P_i(t) / P_C(t)$, where $R^0_C= \sum_{j: D_j=0} R^0_j$ and $P_C(t)= \sum_{j: D_j=0} P_j(t)$.
Defectors retain resource with a penalty, $R_i(t+1 | D_i=1)= R^0_i - w$.
Power changes according to
\begin{eqnarray}
\Delta P_i(t | D_i = 0 ) &=& \Delta t \left[\mu R_i(t) - \mathcal{N}(t)\right] \nonumber\\
\Delta P_i(t | D_i = 1 ) &=& \Delta t \left[ \mu \left(R_i(t)+\rho w\right) - \mathcal{N}(t) \right],
\end{eqnarray}
where $\mathcal{N}$ is a normalising constant to ensure $\sum_i \Delta P_i(t)=1$. In the basic model $R^0_i=1/N$.  Power is initialised by giving each faction half the power of the previous one.

\subsection{Qualitative Indicators}
Section S3 
and Fig. \ref{fig:allsummaries} interpret these scores. The `State formation' score $S_f = 2(Q_1+Q_4)$ where $Q_i$ is the proportion of timesteps where the number of cooperators $C(t)$ is in the $i$-th quartile. The `Periodicity' score uses $p(t) = (T-t)^{-1} \sum_{t^\prime=1}^{T-t}p[C(t^\prime+t)=C(t^\prime)]$ to form $S_p= p(\tau) - [p(\tau/2) +p(3\tau/2)]/2$ where
$\tau = \argmax_{t : t \ge t_{min}} p(t)$, $T$ is the total number of timesteps, and $t_{min}$ excludes the first mode. The `State size' score $S_s =1$ if $\range_t(C(t))=[1,N]$, $S_s =0.5$ if $\min_t(C(t))=1$ and $S_s =0$ otherwise. The `Capital stability' score $S_c=1$ if $\argmax_i(P_i(t))=1$ for all $t$. 

\subsection{Proof of Proposition \ref{prop:gametheory}}
Power during cooperation follows: 
\begin{equation*}
\frac{d P_i(t|\mathbf{D}=\mathbf{1})}{dt} = \mu \left(P_i - \frac{1}{2}\right)
\implies P_i(t) = \frac{1}{2} - \delta \exp(\mu t)
\end{equation*}
starting at $t=0$. The time taken to reach $p_{min}$ is $t_c=(1/\mu) \log(1+a/\delta)$.  The resource obtained per unit time in the cooperation state is $P_i(t)$.  Therefore the total resource obtained is:
\begin{equation*}
R_{c}^T = \int_{0}^{t_c} R_i(t|D_i=0)dt = \frac{1}{\mu} \left[\frac{1}{2}\log\left(1+a/\delta\right)-a\right].
\end{equation*}
During the defection phase, resource is obtained at rate $1/2-w$ and power accrued at rate $(\mu/N) w_o(\rho-1)$, so the defection duration $t_d=aN/[\mu(\rho-1)]$ and therefore $R_{d}^T=\int_{0}^{t_d} R_i(t|D_i=1)dt = a(1/2-w)$.  The average rate of resource acquisition is
\begin{equation}
r_i(a) = \frac{R_c^T+R_d^T}{t_c+t_d} = \frac{\frac{1}{2}\log\left(1+\frac{a}{\delta}\right)-a(1-\frac{\mu}{2})-w\mu}{\log\left(1+\frac{a}{\delta}\right) + \frac{aN}{\rho-1}} \label{eqn:optimalresource}
\end{equation}
This can be solved for $a=a^*$ giving a maximum resource $r^*_{low}$ when the resource rate $\frac{dr(a)}{da}=0$.  Although there is no explicit form, the maxima exists at positive $a$ and is non-trivial (i.e. not a boundary).  
During this time, the leading faction obtains a higher payoff during the cooperation phase $R_j^T=R_i^T+2(\delta+a)/\mu$, with a higher average rate $r_{high}^*$.

\end{materials}


\begin{acknowledgments}
\end{acknowledgments}





\footnotesize
\bibliographystyle{mychicago}
\bibliography{socialcollapse}


\section*{Supplementary material}

# Supporting Information for 'Inequality, unrest and the failure of states: a qualitative modelling framework'

Daniel John Lawson[*] and Neeraj Oak[†]

June 14, 2013

# Contents

# List of Figures

[*]Heilbronn Institute for Mathematical Research, University of Bristol, Bristol, BS8 1TW, UK
[†]Department of Engineering Mathematics, University of Bristol, Bristol, BS8 1TW, UK

# S1 Mathematical frameworks for the basic model

Model 1a removes intrinsic noise (see Section S2.5) by taking the continuous-time limit. Instead of factions being offered a choice to defect according to a schedule, time proceeds until the next change in decision occurs. Multiple decision changes may follow instantaneously if the choice of one faction affects the choice of another. In this way we can understand 'defection cascades' and 'cooperation cascades' in which a large fraction of the actors change their decision at once.

Model 1b allows for a continuous distribution of factions. Because there is no noise in the system, the power ordering of factions cannot change. Therefore a single decision boundary between cooperation and defection will always exist, which may remain constant over time (during a cooperation or a defection phase) or move rapidly (during a cascade). This model takes the form of a Partial Differential Equation.

| Parameter | Explanation |
| --- | --- |
| $N$ | Number of factions |
| $R_i^0$ | Raw resource available to faction $i$ ($R_i^0 = 1/N$ in the basic model). |
| $w$ | Defection 'intensity' controlling resource penalty and power gain |
| $\mu$ | Increment of time. |
| $\rho$ | Power gain during defection relative to $w$. |

Table 1: Description of parameters in the basic model.

Table 1 describes the parameters of the model, which for convenience we reproduce here. The decision to defect $D_i(t) = 1$ if $\eta_i(t) > 0$ and $D_i(t) = 0$ otherwise, where:

$$\eta_i(t) = R_i(t|D_i = 1, \mathbf{D_{-i}}) - R_i(t|D_i = 0, \mathbf{D_{-i}}). \tag{S1}$$

Cooperators pool and redistribute resource, $R_i(t|D_i = 0) = R_C^0 P_i(t)/P_C(t)$ (where $R_C^0 = \sum_{j:D_j=0} R_j^0$ and $P_C(t) = \sum_{j:D_j=0} P_j(t)$), whereas defectors retain resource with a penalty, $R_i(t|D_i = 1) = R_i^0 - w$. Power changes according to

$$\begin{aligned} P_i(t+1|D_i=0) &= P_i(t) + \Delta t \left[\mu R_i(t) - \mathcal{N}(t)\right] \\ P_i(t+1|D_i=1) &= P_i(t) + \Delta t \left[\mu \left(R_i(t) + \rho w\right) - \mathcal{N}(t)\right], \end{aligned} \tag{S2}$$

where $\mathcal{N}$ is a normalising constant to ensure $\sum_i \Delta P_i(t) = 1$. By substituting the normalising factor $\mathcal{N}$ into Eq. S2, defining the number of cooperators $C = N - \sum_{i=1}^N D_i$, the basic model can be written:

$$\begin{aligned} \frac{\Delta P_i(t|D_i=0)}{\Delta t} &= \mu \left[\frac{C}{N}\frac{P_i(t)}{S(t)} - \frac{1}{N} - \frac{N-C}{N}w(\rho-1)\right] \\ \frac{\Delta P_i(t|D_i=1)}{\Delta t} &= \mu \left[\frac{1}{N} + w(\rho-1) - \frac{1}{N} - \frac{N-C}{N}w(\rho-1)\right] \\ &= \mu \frac{C}{N}w(\rho-1) \end{aligned} \tag{S3}$$

where $S(t) = \sum_{i=1}^N (1-D_i)P_i(t)$ is the power held by the cooperators. For convenience we label this as *Model 1*.

*Model 1a*: A continuous-time description of Model 1 can be obtained by taking the limit $\Delta t \to 0$, which we will call Model 1a. The continuous time version of the model is special, because the original power ordering can not change. This can be exploited to obtain several theoretical results which do not exactly hold in the discrete time model, although the general features (such as overall defection patterns) do still hold. We order factions by their power (from high to low); now the number of cooperators $C$ is a defection threshold with $D_i = 0$ for $i \leq C$, $D_i = 1$ for $i > C$. Therefore $S(t) = \sum_{i=1}^{C} P_i(t)$ is the power held by the cooperators, and additionally we can define $S_x(t) = \sum_{i=1}^{x} P_i(t)$ as the power held by all factions up to $x$. The power model becomes:

$$\frac{dP_i(t|i \leq C)}{dt} = \mu \left[ \frac{C}{N} \frac{P_i(t)}{S(t)} - \frac{1}{N} - \frac{N-C}{N} w(\rho - 1) \right]$$
$$\frac{dP_i(t|i > C)}{dt} = \mu \left[ \frac{C}{N} \right] w(\rho - 1) \tag{S4}$$

*Model 1b*: We will make use of a continuous distribution of factions, by taking the continuous limit in faction space $x \in (1, N)$, leading to Model 1b:

$$\frac{\partial P(x,t|x \leq C)}{\partial t} = \mu \left[ \frac{C}{N} \frac{P(x,t)}{S(x,t)} - \frac{1}{N} - \frac{N-C}{N} w(\rho - 1) \right]$$
$$\frac{\partial P(x,t|x > C)}{\partial t} = \mu \frac{C}{N} w(\rho - 1) \tag{S5}$$

where we have defined $S(x,t) = \int_0^x P(x,t)dx$. We force $C \geq 1$ to preserve the special behaviour of the leading faction (now represented continuously by $x < 1$) who always cooperates. Expressed like this the model deals with infinitesimal densities. Since $S(x,t)$ appears on the right hand side, it will be convenient to work with this directly and recover faction power using $P(x,t) = \partial S(x,t)/\partial x$. We will write $S(x|t)$ when considering a fixed $t$ and $S(t|x)$ for a fixed $x$; these equations are follow ODEs rather than PDEs.

Because of the dependence on the defection threshold $C$ (which is defined in terms of the power distribution) searching for a general solution to the ensuing PDE in Model 1b is ambitious. However, we will be able to provide bounds to the distribution $S(x,t)$ that can lead to, and recover from, collapse into a defection state. From this we can bound the time taken for defection periods.

We are interested in the behaviour of all 3 models and will use results from each in the calculations. However, the focus is on properties of Models 1a-b that are also present in the simulations of Model 1.

## S1.1 Behaviour when all factions cooperate

When all factions cooperate, the dynamics of Model 1a are simple. Substituting $D = 0$ into Equation S4 gives $S(t) = 1$ and hence:

$$\frac{dP_i(t)}{dt} = \mu \left[ P_i(t) - \frac{1}{N} \right] \tag{S6}$$

which can be solved by separation of variables, with initial condition $P_i(t=0) = P_i^0$ to give:

$$P_i(t) = \frac{1}{N} + \left( P_i^0 - \frac{1}{N} \right) \exp(\mu t) \tag{S7}$$

i.e. exponential departure from the initial deviation away from the mean initial resource (which was $1/N$). We can therefore solve for $t_i^D$, the time until faction $i$ defects. This occurs when $R_i(t) = 1/N - w$, and as additionally we have $P_i(t) = R_i(t)$, we obtain:

$$t_i^D = \frac{1}{\mu} \log\left(\frac{w}{P_i^0 - 1/N}\right) \tag{S8}$$

Only the minimum time (hence $\arg\min_i(P_i^0) = P_N^0$) is relevant to the time of the first defection event.

## S1.2 Defection cascades

A defection cascade is defined by defections occurring as a direct result of other defections. Faction $N$ is first to defect and has $P_N(t) = R_N(t) = 1/N - w$; subsequent defectors have greater power.

Defecting factions have resource $R_i(t|D_i = 1) = 1/N - w$, hence the decision for faction $C$ to defect requires:

$$\frac{C}{N} \frac{P_C(t)}{\sum_{j=1}^{C} P_j(t)} - \frac{1}{N} + p_w < 0 \tag{S9}$$

which can be solved for the lower bound $P_C(t) < b_C$ for the faction $C$ to defect (recalling that faction $C = 1$ cannot defect):

$$b_C = \frac{1 - wN}{(C-1) + wN} \sum_{j=1}^{C-1} P_j(t) \tag{S10}$$

Note that $b_{C-1} \geq b_C$ since $P_j$ is not decreasing with $j$.

We now work with the cumulative power $S_i(t) = \sum_{j=1}^{i} P_j(t)$, and solve for the case $b_C = P_C(t) = \Delta_i S_i(t)$ using the notation $S_i^*(t)$ for the cumulative power up to the boundary. Letting $P_C(t) = b_C$ acts as an effective upper bound for defection of faction $C$, since it places the minimum possible power with the remaining cooperators, making defection for one of them harder.

Recall that cooperators have small $i$, so $S^*$ is the power in the remaining cooperators; defection of faction $i$ will occur if the faction is too weak to remain with the other cooperators, i.e. $S_i^*(t)$ is too large. Rearranging Equation S9 for the boundary:

$$S_{i+1}^*(t) - S_i^*(t) = \frac{1 - wN}{i + wN} S_i^*(t) \tag{S11}$$

which when considered in Model 1b leads to

$$\frac{dS^*(x|t)}{dx} = \frac{1 - wN}{x + wN} S^*(x|t). \tag{S12}$$

Note that this restricts $0 < w < 1/N$. Equation S12 can be solved by separation of variables (recalling that $S(x = N|t) = 1$):

$$S^*(x|t) = \left(\frac{x/N + w}{1 + w}\right)^{1 - wN}. \tag{S13}$$

4

Instead of differentiating $S^*(x|t)$, which would approximate $S_i^*(t)$ via a piecewise linear function, we will obtain the best match with Models 1 and 1a if we compute the boundary for specific factions using differences of the cumulative power of factions:

$$P_i^*(t) = S_i^*(t) - S_{i-1}^*(t) = \left(\frac{i/N + w}{1+w}\right)^{1-wN} - \left(\frac{(i-1)/N + w}{1+w}\right)^{1-wN} \tag{S14}$$

For a defection cascade to occur in Model 1b, $S(x|t) > S^*(x,t)$ is a strict requirement on the distribution of power – any faction reducing the total cooperation power below this has sufficient power to continue cooperating. A defection cascade will stop at the first faction for which the bound does not hold. However, it is possible (and occurs in practice) that defection cascades instantaneously reach only a large proportion of the factions. The remaining few factions have power above the threshold, but their power continues to reduce, so more may join the defection. If $N$ is large, several factions may avoid defection completely.

Defection cascades can occur instantaneously in Models 1a and 1b. For Model 1, assuming $\mu$ is small, a defection cascade changes the resource levels by a much larger amount than it would change under normal time evolution of the model. Therefore cascades in Model 1 (and Model 1a) take a very similar form, differing only qualitatively.

## S1.3 Duration of epochs in the cycle

Collapse (and recovery) events are triggered by the whole cumulative power distribution falling above (below) the decision threshold. This is a complex procedure in general as smaller defection and cooperation events affecting a few factions can occur, which can change the shape of the resulting distribution.

We can obtain a bound on the maximum duration of cooperation and defection period. We assume that all factions act at the same time, do not defect independently during the cooperation phase, and do not cooperate independently during the defection phase. This implies that all defecting factions have the same power, as they experience an identical increase in power during the defection period, and therefore must experience the same decrease during the cooperation period.

This model maximises the time spent in a defection by restricting the first cooperation event. Consequently, the leading faction has been reduced to the minimum power possible and so the cooperation period starts from the fairest distribution of power possible. This maximises the time spent in the cooperation phase.

Collapse occurs when faction $N$ defects, and recovery occurs when faction $N$ can resume cooperation (as $P_i(t) = P_N(t)$ for all $i \geq 2$). As $(N-1)P_N(t) + P_1(t) = 1$, we can describe the dynamics in terms of a single quantity $P_N(t)$. Let $\tau_c$ be the time for collapse, $\tau_r$ be the time for recovery. Let $P_N(\tau_c) = P_N^c$ be the power of faction $N$ at the collapse event, and $P_N(\tau_r) = P_N^r$ be the power at the recovery event. The duration of the cooperation phase is $t_a = \tau_c$ and the duration of the defection phase is $t_b = \tau_r - \tau_c$.

The condition for defection gives $P_N^c = 1/N - w$, and the condition for recovery (from Equation S10) gives $P_N^r = \frac{1-wN}{1+wN}P_1^r$ which (since power sums to 1) implies $P_N^r =$

5

$\frac{1}{N} \frac{1-wN}{1-w(N-2)}$. During the cooperation phase (from time 0 to $t_c$, duration $t_a$)

$$\frac{dP_i(t|C=N)}{dt} = \mu \left( P_i(t) - \frac{1}{N} \right)$$
$$\implies P_N^c = \frac{1}{N} + \left( P_N^r - \frac{1}{N} \right) \exp(\mu t_a) \tag{S15}$$

leading to

$$t_a = \frac{1}{\mu} \log \left[ N \left( w + \frac{1 - Nw}{2} \right) \right]. \tag{S16}$$

Similarly, during the defection phase (from time $t_c$ to $t_r$, duration $t_b$)

$$\frac{dP_N(t|C=1)}{dt} = \mu \frac{1}{N} w(\rho - 1)$$
$$\implies P_N^r = P_N^c + t_b \mu \frac{1}{N} w(\rho - 1) \tag{S17}$$

leading to

$$t_b = \frac{1}{\mu(\rho-1)} \frac{(N-2)(1-Nw)}{1-w(N-2)}. \tag{S18}$$

This bound is extremely accurate when initial conditions are close to the assumptions (i.e. factions $2-N$ have equal power). Under some parameter and initial conditions, the distribution of power tends towards this structure, in which case the bounds are moderately close (but not exact due to intrinsic noise). If factions have a distribution broad distribution of power in stationarity, then the bound is poor. A significant factor is that defection and cooperation cascades do not occur instantaneously, and if multiple factions retain power above $1/N$ then the cooperation cascade is structured differently.

## S1.4 Informative reformulations

When all $R_i^0 = R^0$ the resource defection penalty can trivially be written as a relative reduction in resource, rather than a subtraction:

$$R_i(t|D_i = 1) = R^0 - w = aR^0 \tag{S19}$$

where $a = 1 - w/R^0$. Similarly, the political gain for rebellion can be rewritten as a relative rather than absolute gain:

$$P_i(t+1|D_i = 1) = P_i(t) + \Delta(t) \frac{\mu}{N} [R_i(t) + \rho w] = P_i(t) + \Delta(t) \frac{\mu \lambda}{N} R_i(t) \tag{S20}$$

where $\lambda = 1 + \rho w/(R^0 - w)$. This form makes it clear that power under both choices are exponentially changing with different rates. The model has not changed under these reformulations.

6

## S2 Model Extensions

### S2.1 Unequal resource distribution

Most of the theory we have developed applies only to the case where all factions have equal access to resources. When these differ, each will have a different decision boundary and obey more complex dynamics (as e.g. when cooperating they will be diverging from different points). Although the mathematical details do differ, the intuitive reasoning remains the same: each faction has a decision boundary, and changes of decision will still have a cumulative effect on other factions' decision boundaries. Therefore (provided all factions can actually afford to defect) the qualitative dynamics remain the same.

In Figure 3e we use a parametric model for the unfairness of the resource distribution, parameterised by the inequality parameter $\kappa$:

$$R_j^0 = \frac{\exp(\kappa j)}{\sum_{i=1}^{N} \exp(\kappa i)} \tag{S21}$$

Figure S1 shows the structure of this model as the intrinsic resource available to factions, and the initial conditions, are varied. When the initial distribution of power is correlated with resource (Figure S1A), then the dynamics of the model are almost unchanged from the uniform resource model. However, political leadership can be taken from the resource-weak by a resource-strong faction (Figure S1B). It is possible for an faction to remain leader when they only have a moderate resource (Figure S1C).

Since the most powerful faction is $j = 1$ and the most resource-rich faction for $\kappa > 0$ is $j = N$, there is a conflict between power and raw resource. Hence Figure 3e involves a reversal of the capital for a wide range of $\kappa$. However, for smaller $\kappa$ the reversal does not occur and a resource-poor faction remains in power. Figure S1 illustrates how this results in *lowered* overall conflict rate, because average inequality remains at a lower level. When resource rich factions are in power (Figure S1AB, other factions much defect frequently to maintain their position. When resource poor factions are in power (Figure S1c), resource-rich factions do not need to defect much, and other resource-poor factions are unaffected, resulting in a lower defection rate.

### S2.2 Uncertain outcomes

Power has been modelled using deterministic dynamics conditional on the defection choices. We relax this unrealistic assumption by allowing noise in the outcome of the political power process. If we let $\Delta t P_i^D(t)$ be the deterministic prediction for the change in power from Equation S2, then the power follows

$$P_i(t) = \Delta t P_i^D(t) + \Delta t N(0, \sigma_p), \tag{S22}$$

which in the continuous time limit (cf Model 1a) can be seen as an Ornstein-Uhlenbeck (OU) process. This process is 'mean reverting' to the deterministic solution but does deviate on average by $\sigma_p/2\mu$ (by rescaling and taking the continuous limit).

The appropriate scale for the noise is the difference between the power of the factions. From Section S1 this is $O(1/N)$ so $\sigma_p/2\mu \ll 1/N$ for the deterministic dynamics to dominate. Figure 3f and Figure S2 support this. Noting that we used $\mu = 0.05$ and $N = 11$ here leading to $\sigma_p \ll 0.009$, the asymptotics have worked surprisingly well. Increasing

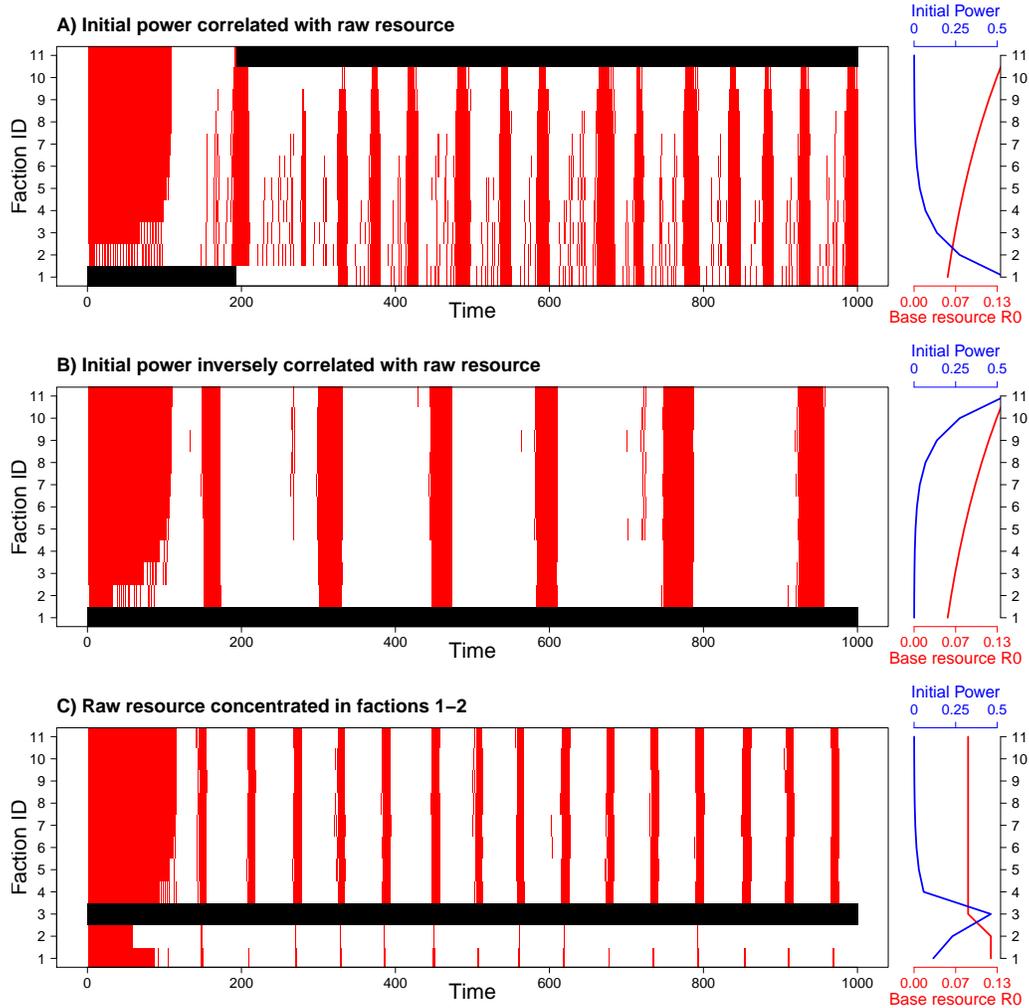

Figure S1: Effect of changing raw resource levels. The left plot shows the defection behaviour as in Figure 1C, and the right plot shows the initial power distribution (blue line, top axis) and base resource level $R_i^0$ (red line, bottom axis). Shown are A) $\tau = 0.1$, B) $\tau = -0.1$, and C) initial power is distributed such that the highest resource factions are not leaders.

$\sigma_p$, leader replacement occurs first, then periodicity is lost, followed by infrequent state formation and finally a failure to coordinate collapse events.

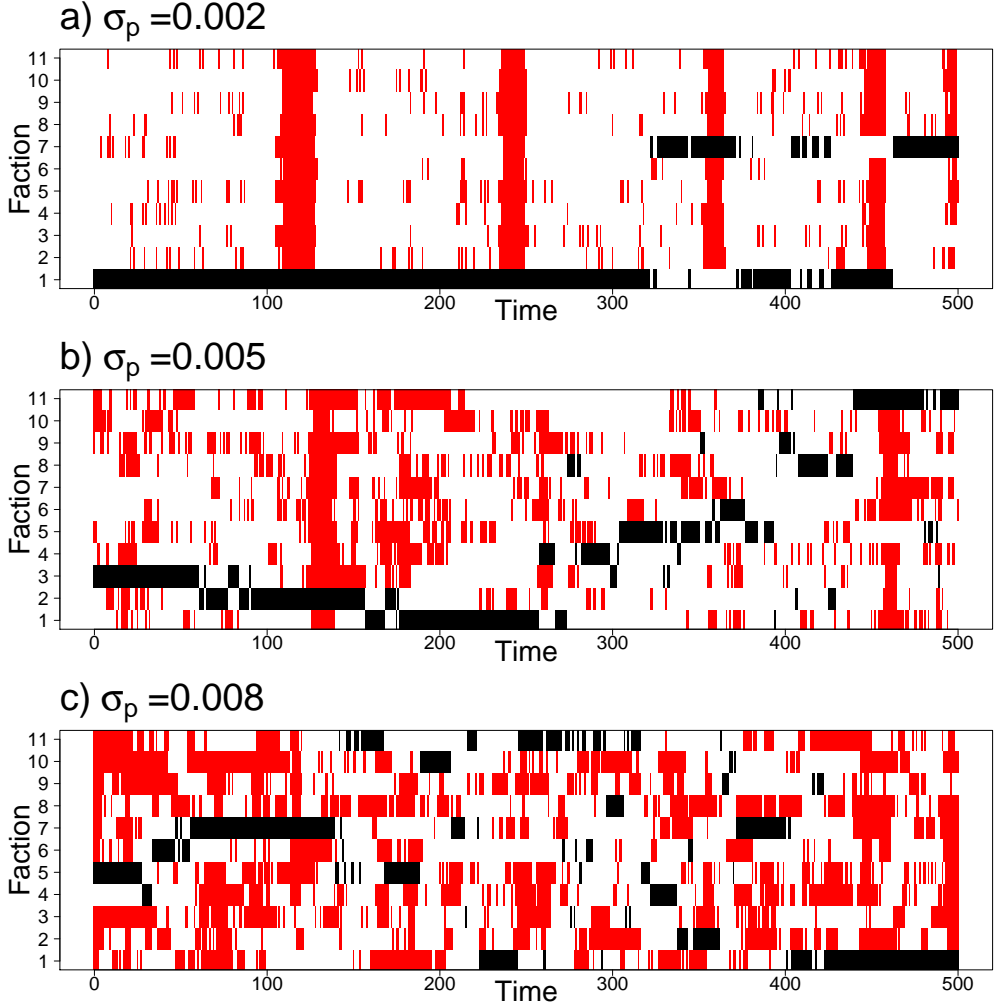

Figure S2: Effect of changing the randomness in power outcomes $\sigma_p$. a) $\sigma_p = 0.002$, for which leader replacement is rare but possible. b) $\sigma_p = 0.005$, for which periodicity has been broken down but collapse events persist. c) $\sigma_p = 0.008$ for which noise is significant and state formation is rare (but possible), and collapse events are poorly coordinated.

### S2.3 Biased decision making and random choices

To generate an 'imperfect decisoon model' we replace the short-term resource optimising decision rule for $\eta$ with a random function taking the form

$$\eta_i(t;\tau,\sigma,\beta) = R_i(t|D_i=1) - R_i(t|D_i=0) + \beta[2D_i(t-1) - 1] + \sigma G(t;\tau), \qquad (S23)$$

9

where $\beta$ is the bias towards the previous choice. $G \sim \text{GP}(\tau)$ is a Gaussian Process, which is a random function with expected mean 0 and standard deviation 1. Specifically, for a specified set of times $t_1 \cdots t_N$, we define $G(t_1, \cdots, t_N) \sim \text{MVN}(0, K)$ i.e. multivariate normally distributed with covariance matrix $K_{i,j} = \tau^2 |t_i - t_j|^2$. One of the reasons to use a Gaussian Process is that they defined for all times but only need to be evaluated at the times where they are needed.

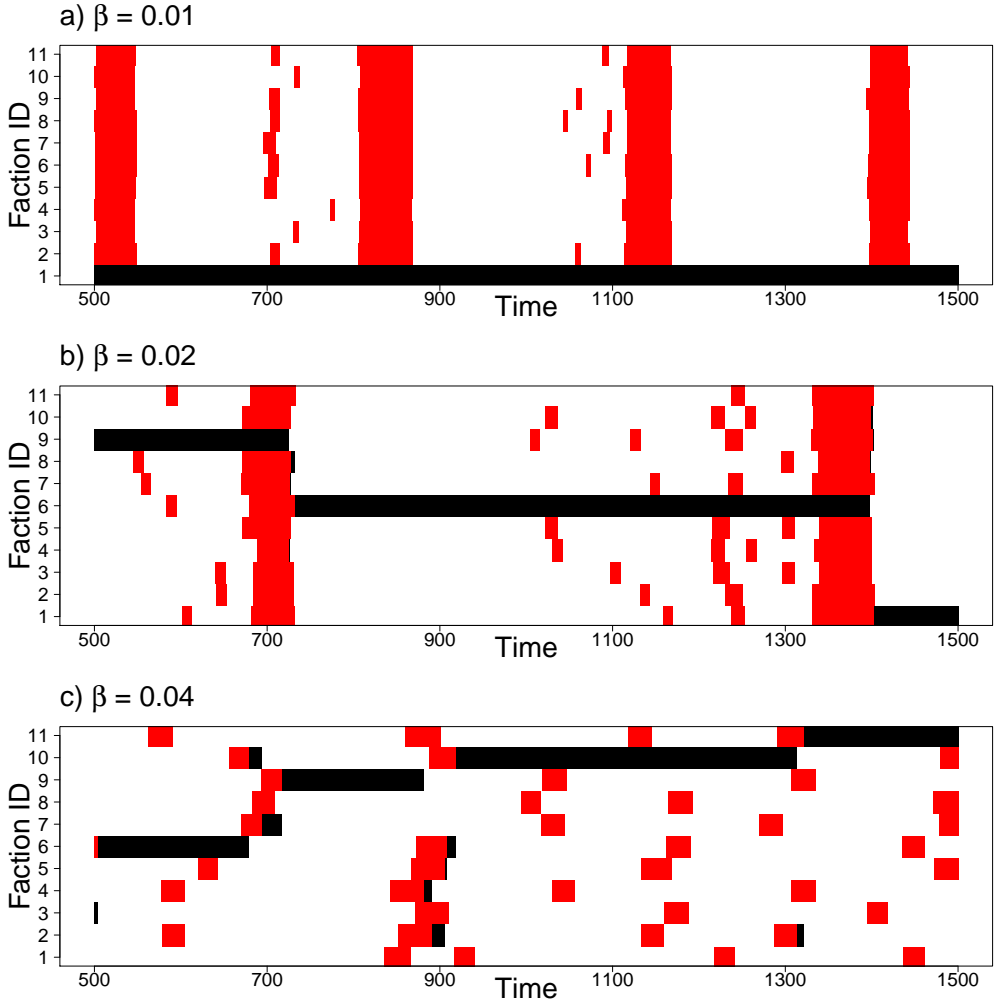

Figure S3: Behaviour of the Imperfect decision model with persistent choices, varying $\beta$. a) Small $\beta = 0.01$ produces behaviour that is qualitatively the same as $\beta = 0$, with a slightly longer period. b) Intermediate $\beta = 0.02$ allows leader turnover and increases the period further. c) Large $\beta = 0.04$ prevents full scale collapse events but increases the rate of collapse events, increasing the leader turnover rate. As before, $w = 0.02$, $\rho = 20$, $\mu = 0.01$ and $N = 11$.

Figure S3 illustrates the effect that a tendency to keep the same decision (parameterised by $\beta$) has on the behaviour of the model. In general, it produces longer periods of both defection and cooperation, as factions are reluctant to change. However, this has the effect of forcing a more fair initial distribution of power in the forming state, which permits the state to survive for still longer. Defections for a single faction have a defined minimum length, since it takes a significant improvement of power to reverse a decision. When $\beta$ is moderate (Figure S3b), repeated defection can completely erode the leaders resource benefit, resulting in leader turnover. For very large $\beta$ (Figure S3c) the behaviour of others does not affect choices very strongly compared to the desire to maintain the previous action. In this case defection cascades are partially prevented.

The model with $\tau = 0$ has independent Gaussian noise which is faster to simulate from. The tendency not to change decisions can be easily be motivated from a psychological perspective, or using internal politics if factions are groups of individuals. For example, a faction leader may make a promise to cooperate or defect, which is difficult to reverse without losing face. It can also be seen as risk aversion for the unknown effects of new actions – factions may find it difficult to predict whether others will defect if they do.

Figure S4 demonstrates how the model breaks down in the presence of noise. Small $\sigma$ relative to the intrinsic power variability for non-leading factions has little effect on the dynamics. However, as $\sigma$ increases it becomes more likely that factions will defect during stable cooperation. Moderate $\sigma$ leads to a reduction in the length of collapse periods (Figure S4b). High $\sigma$ lets factions make choices 'randomly' (Figure S4c), which prevents the possibility of state formation as no faction can remain in power.

However, correlated decisions can potentially permit cooperation in the presence of very large decision noise. Figure S5a-b shows the behaviour in the intermediate noise case with $\sigma = 0.012$. With $\tau = 5$ (Figure S5a) cooperation periods have markedly increased average cooperation levels from the $\tau = 0$ case (Figure S4b), cooperation periods are shorter and collapse times are less predictable. Increasing $\tau$ to 25 (Figure S5b) restores the very clear distinction between cooperation and defection periods. This persists even at high $\sigma$ (Figure S5c), with the effect that leader turnover is now possible as factions may refuse to cooperate for long enough to remove the leaders natural resource advantage.

The reason that highly correlated noise can support cooperation and defection periods in our model is that each faction has a new equilibrium resource level that they will change their action at (in the short term). If this changes more slowly than the internal political dynamics, then they will change behaviour due to the effect of others rather than due to their own random choices.

## S2.4 Spatial structure

Some political scenarios are best described with a spatial model. For example, factions may be local leaders of villages, or semi-autonomous regions of a larger state. Spatial structure can be introduced via the relative advantage that cooperation confers. It is natural that both the resource penalty for defection, and the political gains from doing so, should depend on physical location. We can achieve this by replacing $w$ by $w_i$, which is a function of distance from the political leader. Specifically, we allow an exponential decay in the impacts of defection with distance:

$$w_i = w_0^* \exp\left(-|x_i - x_C| w_d / N\right) \tag{S24}$$

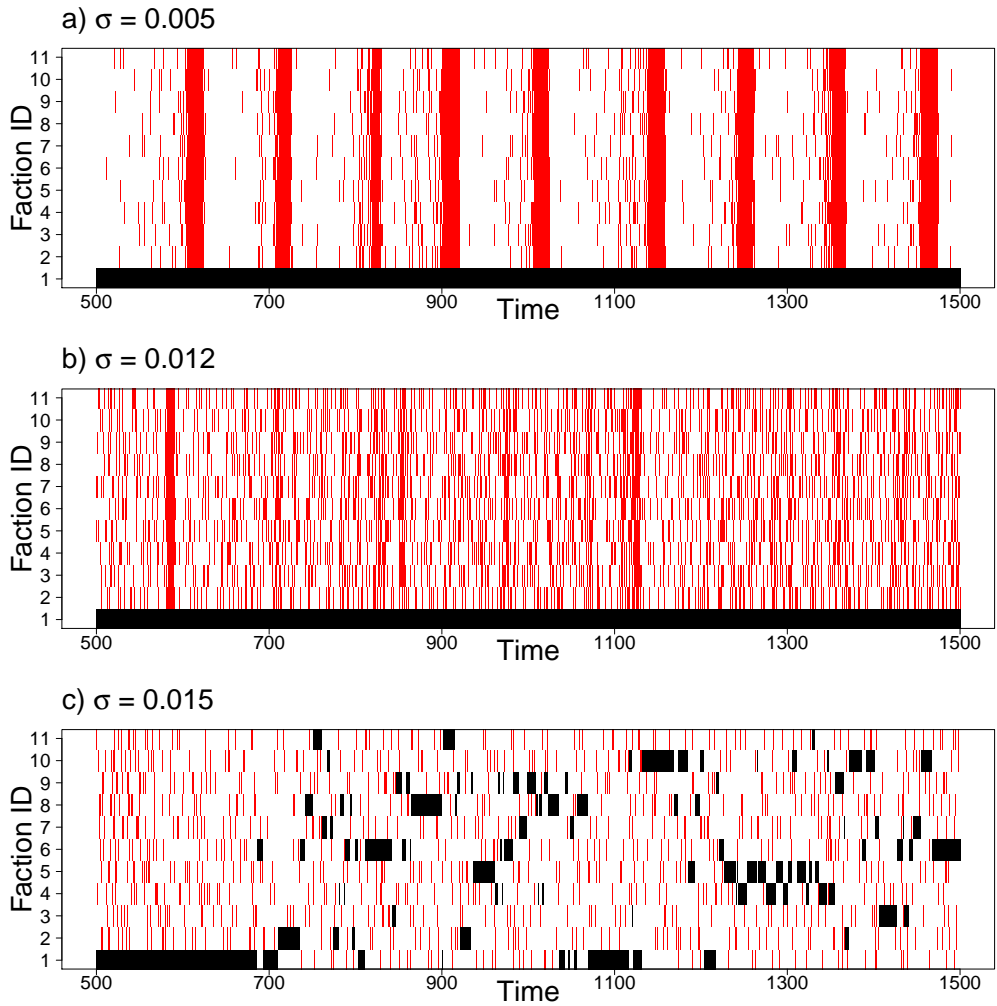

Figure S4: Behaviour of the imperfect decision model with random choices, varying $\sigma$ but fixed $\tau = 0$. In a) low noise levels lead to the same qualitative dynamics as in the no-noise case, in b) medium noise levels reduce coordination of collapse, leading to less define collapse events. In c) high noise levels prevent the construction of a permanent 'state', with turnover of the leading faction possible but state formation involving all factions is not possible. As before, $w = 0.02$, $\rho = 20$, $\mu = 0.01$ and $N = 11$.

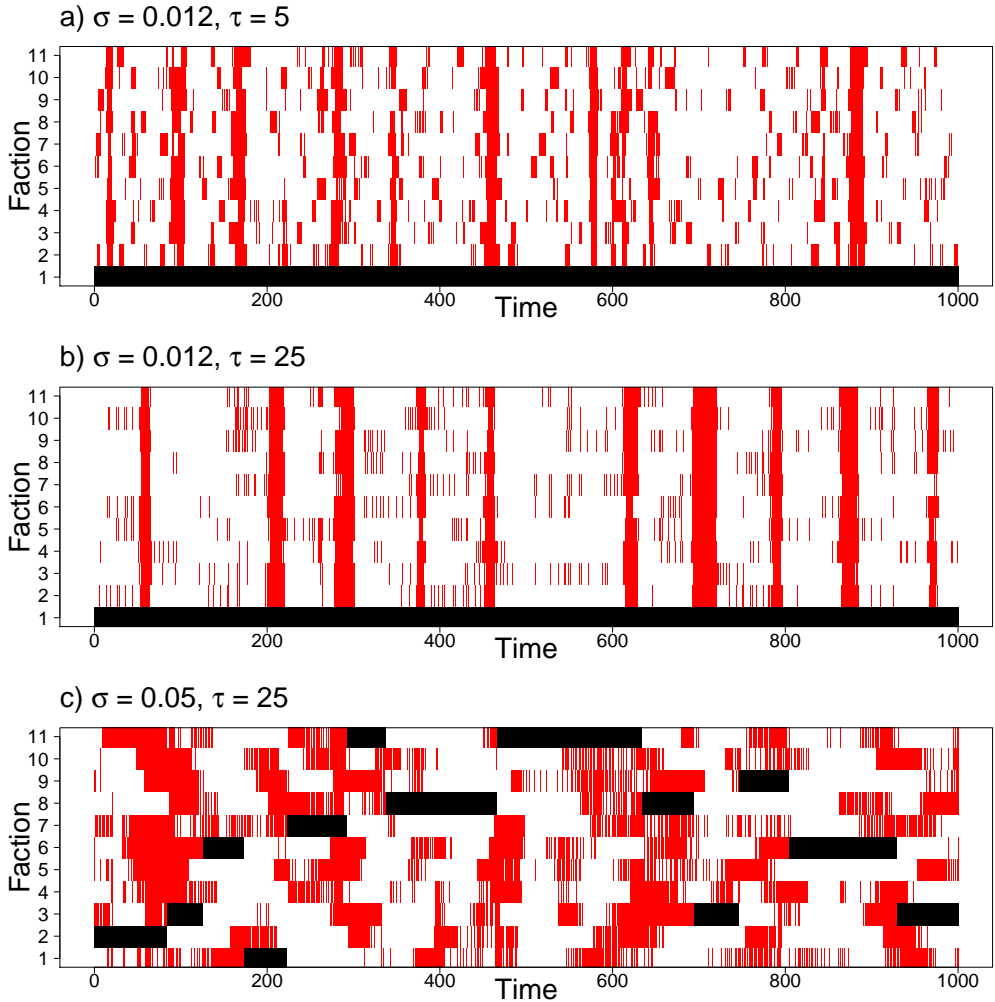

Figure S5: Behaviour of the imperfect decision model with random choices, varying $\sigma$ and decision correlation time $\tau \neq 0$. a) and b) have $\sigma$ comparable to Figure S4b, but vary the decision correlation time $\tau$; c) is comparable with Figure S4c with large $\tau$. As before, $w = 0.02$, $\rho = 20$, $\mu = 0.01$ and $N = 11$.

where $w_0^*$ is a normalising factor to ensure the average effect is the same as the non-spatial model, $E(w_i) = w$, and $w_d$ is the spatial decay rate. When $w_d = 0$ the model reduces to the basic model. The factor $N$ is included to account for the size of the system, as $x_i = i$ is the spatial location of each faction (we consider them on a ring so that faction 1 is next to faction $N$, to remove boundary effects).

The spatial model in Figure S6 allows for a variety of different scenarios. Interpreting cooperation as a political state, this will grow from the 'capital' (the leading faction) and will collapse as in the non-spatial model. Collapse may be from the outside-in (Figure S6b)

13

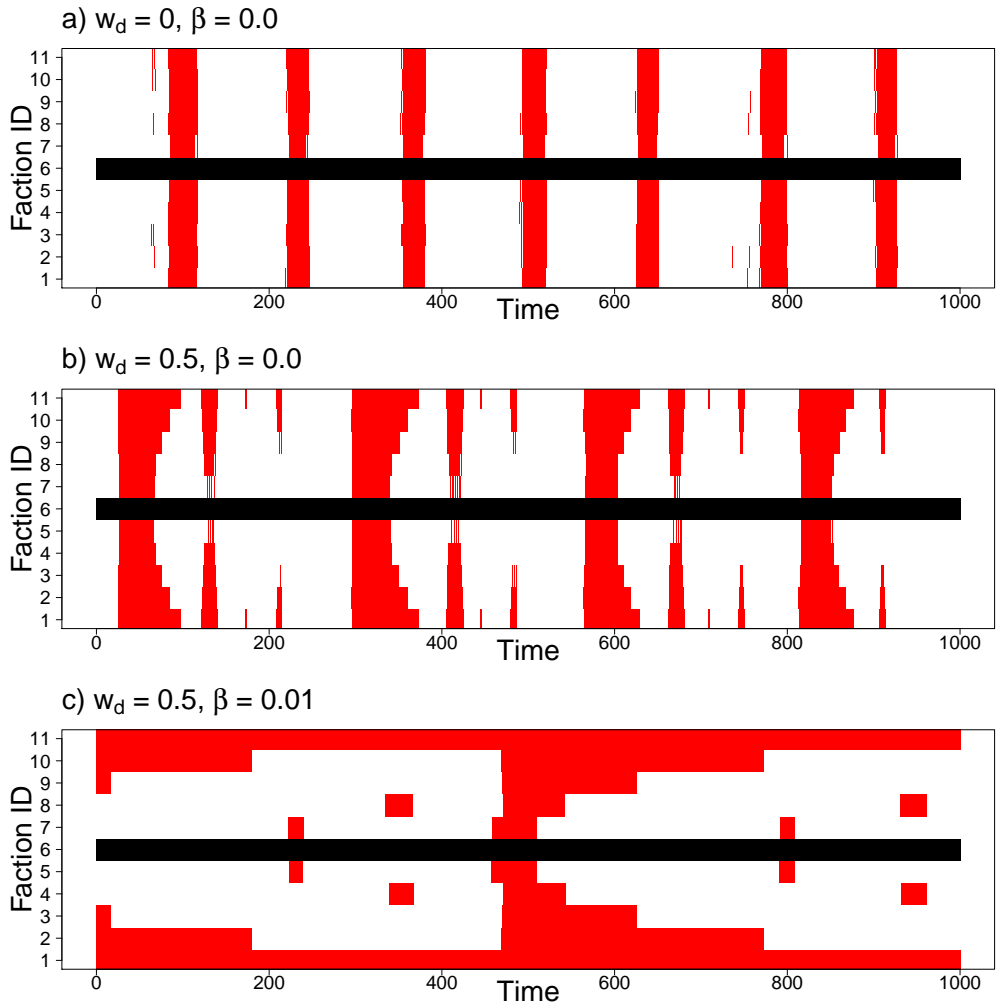

Figure S6: Behaviour of the spatial model (Model 3) with different values of the spatial structure strength $w_d$. a) $w_d = 0$ and $\beta = 0$, for comparison. b) $w_d = 0.5$ and $\beta = 0$. c) $w_d = 0.5$ and $\beta = 0.01$. As before, $w = 0.02$, $\rho = 20$ and $N = 11$, but we use $\mu = 0.005$ for greater temporal resolution of the collapse events.

or the inside-out (Figure S6c). There may be a well defined maximum spatial extent. As in the non-spatial model, the leader will remain stationary unless the tendency to repeat the last action ($\beta$) is high enough. Because the change in power is highly dependent on location, the duration of cooperation and defection periods changes drastically, but the qualitative features of the non-spatial model remain.

14

## S2.5 Modified intrinsic noise

We chose to define the basic model as an iterated game. However, this has consequences for the way that noise enters the system. There is nothing stochastic in the definition of the model, but the discretisation of time can produce 'chaotic' dynamics as small variations in the value of political power have large effects if they interact with the decision boundary between defection and cooperation. Additionally, power changes relatively rapidly during defection, which means that non-linear effects of the interaction of various states will make the continuous time version of the model behave differently.

Intrinsic noise arises from the discrete nature of the choices and the times at which those are made. If the power of state $i$ changes by $\delta P_i(t)$ and it has a decision boundary at $P_i^*$ then there are a range of power values that lead to the same outcome of a change in decision. Since power changes non-linearly, the particular values of power can get out of phase, but can also be reset into phase during coordinated defection by changing the order of cooperation. This can lead to semi-regular noise structures, such as that observed in Figure 1. From Equation S3 the noise is of intrinsic magnitude $O(\mu/N)$ (with situation-specific dependence on the other parameters).

There are three possible views of intrinsic noise:

1. Treat the underlying continuous-time, noise-free model as the model of interest (Model 1a, described by Equation S3). Intrinsic noise is viewed as a numerical integration error, and alternative numerical schemes could be considered.

2. Treat $\mu$ as a real parameter in the model, and treat the chaotic dynamics as a 'real' source of noise.

3. Consider alternative models that characterise the discretisation differently, and focus on properties of the system that are common to all models.

Although it would be interesting to study, we have not performed numerical integration. This is because a) a somewhat specialist approach would be required to handle the choice dynamics, and b) we do not believe that the noiseless version of the model is 'closer to reality' than the discretised versions. We also don't treat $\mu$ as an important parameter, again because we believe that extrinsic noise will be present in real-world examples. We instead focus on features that are robust to the level and type of intrinsic noise. To this end we construct an additional variation of the basic model.

*Model 1c*: rather than updating *all* factions at every timestep, we increment time by $\delta t \sim \exp(1/N)$ (so that $E(\delta t) = 1/N$) and update a single, *random* faction. We note that this takes the form of the Gillespie Algorithm and is appropriate if factions are independently updating their decisions at rate 1. (Since the number of factions is constant in our model, there is no interesting difference between this model and that with $\delta t = 1/N$ exactly). Time increments by 1 on average after $N$ iterations of the new timestep:

1. A random faction $i$ can choose whether to cooperate (determining $D_i(t)$ from $\eta_i(t)$). All other factions perform their previous action.

2. The actual resource obtained is evaluated.

3. The corresponding power changes are computed with Equation S2.

15

This is slightly preferable to the original definition as we have removed one cause of chaotic dynamics (the simultaneous decision problem). This prevents 'alternating' where two sets of factions are exactly out of phase with one another (seen in Figure 1a).However, this algorithm takes $O(N)$ more computing power, contains an explicit form of noise, and does not produce any qualitative difference in behaviour.

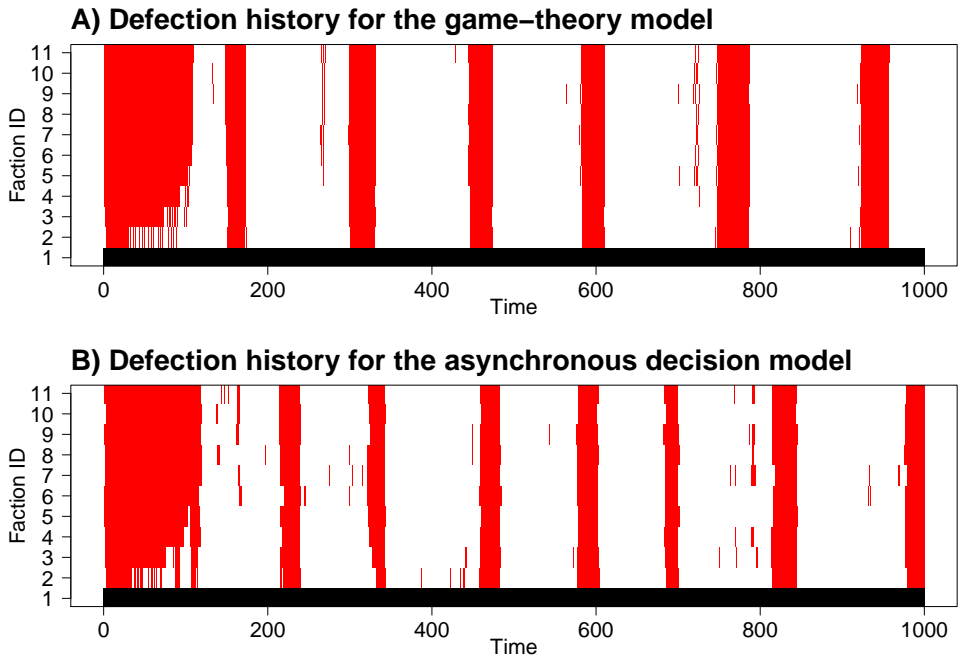

Figure S7: The modified intrinsic noise model, comparing a) the standard synchronised decision model, and b) the asynchronous decision model. Asynchronous decisions do not change the qualitative periodic dynamics, but the period has changed slightly.

Figure S7 shows both the synchronous and asynchronous models, which display the same overall features of periodicity formed by collapse/recovery cycles. The nature of the intrinsic noise has changed; although the discretisation noise is $O(N)$ smaller, there is additional noise in the random decision order. This leads to a change how defection occurs in the cooperation phase - it is more common, less coordinated, and (unlike the synchronous case) does not involve the same ordering of factions. A consequence is that both defection periods and cooperation periods are slightly shorter.

There is not a trivial reparameterisation that leads to the same periodicity. However, a parameter mapping exists to retain the same periodic structure, and the models will only be distinguishable by their fine-scale structure.

### S2.6 Non-uniform defection penalty

We might believe that the magnitude of the penalty for defection should depend on the number of cooperators, such that few defectors get a larger relative resource penalty than

16

when there are many defectors. This would also eliminate the somewhat artificial cooperation benefit enjoyed by the political leader even when there are no other cooperators. Most reasonable models will make it easier for full cooperation and defection cycles to occur. This is because defecting is more difficult when there are few defectors, and cooperating is more difficult when there are few cooperators. Provided that decision changes are still possible for some achievable political power, then cycles will still occur and the qualitative dynamics will be as the basic model. Hence we have focussed on the uniform defection penalty model as it is the 'hardest' model we considered for producing collapse events.

We substitute a variable resource penalty for $w$ into the basic model:

$$w_i(t|S_i = 1, \{S_j\}) = \frac{Nw}{N-2}\left[\left(\sum_{j;S_j=0} R_j^0\right) - R_i^0\right] = w\frac{N-D-1}{N-2}. \qquad \text{(S25)}$$

Here, when 'all defect' (except the leader) $D = N - 1$ the penalty is $w_i = 0$; when all cooperate, the penalty for the first defection (with $D = 1$) is $w_i = w$.

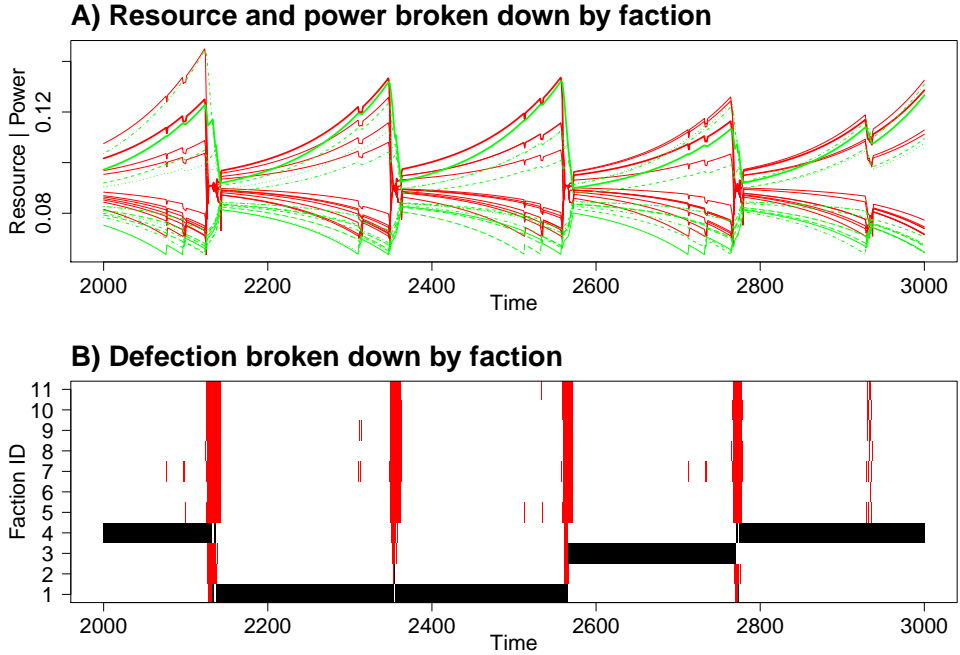

Figure S8: Plot of the dynamics in the variable resource penalty model, with penalty for defection proportional to the number of cooperators. A) shows the resource (red) and power (green), whereas B) shows the defection history.

Figure S8 shows the behaviour of this model, which allows for leader turnover in a nontrivial way. In simulations, we find that the first few factions (in initial political power) all manage to lead cooperation phases, but that politically poor initial circumstances prevent

17

ever becoming the leader. This dynamic occurs because the decision boundary during defection phases has moved significantly higher. A defector considering joining a single cooperator has a boundary with $w$ replaced by $w/(N-2)$, and therefore the boundary is an order of $N$ closer to the leader's. Intrinsic or extrinsic noise is much more likely to lead to a replacement event. However, only factions who retain above-average power can participate in this lead, as others are marginalised during the cooperation phase.

### S2.7 Non-linear relationship between power and resource

Power changes might not scale linearly with resource. We can consider a modification of Equation S2 in which power changes at some power of $R_i$ (which can depend on the defection status):

$$\begin{aligned} P_i(t+1|D_i=0) &= P_i(t) + (\mu/N)R_i(t)^\alpha - \mathcal{N}(t) \\ P_i(t+1|D_i=1) &= P_i(t) + (\mu/N)[R_i(t) + \rho w]^\beta - \mathcal{N}(t) \end{aligned} \quad (S26)$$

where $\alpha$ and $\beta$ take values in $(0, \infty)$ and $\mathcal{N}(t)$ is redefined to maintain a total power of 1. A trivial examination of the continuous time limit of this model (as Model 1a) makes it clear that this simply affects the rate at which power exponentially departs from $1/N$ and therefore cannot have an important consequence for the modelling.

An additional model that might be considered allows resource to be non-linear in power within the cooperators. This again cannot have an important impact as it only changes the decision boundary, and can otherwise be written in terms of Model 1b with $\beta = 1$ and $\alpha \neq 1$.

## S3 Qualitative indicators

Figure 3 quantified the match of our model to the qualitative data by use of four indicators. Although crude, we have checked in all shown cases that these have matched our intuitive understanding of how our model was intended to do. Here we expand more precisely on these measures, which all take as input the second half of a long run (10000 time steps) to ensure that we are observing the long-run behaviour.

1. 'State formation' ($S_f$), measured by the proportion of time the number of conforming or defecting factions is very high or very low. A condition for a 'state' in our model is that it retains a high faction of the factions for a long duration. Specifically: we tabulate the number of timesteps the each number of factions $(1, 2, \cdots, N)$ defects. These are split into quartiles $Q_{1\ldots 4}$, accounting for the uneven number of classes found in each quantile by allowing boundary cases to contribute to both quartiles weighed to ensure that all quartiles would have the same value under a uniform distribution. The score $S_f = 2(Q_1 + Q_4)$.

2. 'Periodicity' ($S_p$), measured by how predictable conforming decisions are in $T$ timesteps. Specifically, we form the probability of a match $t$ timesteps apart, $p(t) = (T-t)^{-1} \sum_{t'=1}^{T-t} p(C(t'+t) = c(t'))$ for the number of conforming factions $C(t)$. We find the time difference $\tau = \arg\max_t p(t)$ for which $p(t)$ is maximal (excluding the first mode at $T = 0$), and then report the height of the probability peak as the score $S_p = p(\tau) - [p(\tau/2) + p(3\tau/2)]/2$ (which should compare the height of

18

the largest peak to the height of the 'troughs' each side of it). This produces similar estimates of $\tau$ to the (more standard) autocorrelation function but in practice resulted in slightly smoother estimates of $\tau$ and $S_p$.

3. 'State size' ($S_s$), scored as 1 if all factions have simultaneously cooperated and simultaneously defected, 0.5 if they all simultaneously defected, and 0 otherwise. This allows us to detect collapse either in full-sized, or smaller states.

4. 'Capital stability' ($S_c$) is 1 only if the leading faction does not change from the initially most powerful faction, and $S_c = 0$ if it has done at any point in the history.

| Score | Match | Deviate | Fail |
|---|---|---|---|
| $S_f$ | $> 0.5$ | NA | $\leq 0.5$ |
| $S_p$ | $> 0.05$ | $\leq 0.05$ | NA |
| $S_s$ | 1 | 0.5 | 0 |
| $S_c$ | 1 | 0 | NA |

Table 2: How scores qualitatively match the data. For a simulation to be classified as a 'match' all individual scores must match. To be classified as 'deviation' one or more individuals scores must deviate and all others match. If any scores fail then the model is classified as 'fail'.


\end{document}